\documentclass[12pt]{article}
\usepackage[dvips]{epsfig}
\usepackage[dvips]{graphics}
\usepackage{a4}
\usepackage{amssymb}

\begin{document}
\setlength{\unitlength}{1mm}

\title{On the origin of divergences in massless $QED_2$}
\author{Rodolfo Casana$^{1\footnote{casana@cbpf.br}}$,
Sebasti\~ao A. Dias$^{1,2\footnote{tiao@cbpf.br}}$\\
 \it{\small $^1$Centro Brasileiro de Pesquisas F\'{\i}sicas}\\
\it{ \small Departamento de Campos e Part\'{\i}culas}\\ \it{\small
Rua Xavier Sigaud, 150, 22290-180, Rio de Janeiro, Brazil}\\
\it{\small $^2$Pontif\'\i cia Universidade Cat\'olica do Rio de
Janeiro}\\ {\it \small Departamento de F\'\i sica}\\ {\small \it
Rua Marqu\^es de S\~ao Vicente, 225, 22543-900, Rio de Janeiro,
Brazil}}
\date{ }
\maketitle


\begin{abstract}
We show that ultraviolet divergences found in fermionic Green's
functions of massless $QED_2$ have an essentially non-perturbative
nature. We investigate their origin both in gauge invariant
formalism (the one where we introduce Wess-Zumino fields to
restore quantum gauge invariance) and in gauge non-invariant
formalism, mapping two different but equivalent mechanisms
responsible for their appearance. We find the same results in both
approaches, what contradicts a previous work of Jian-Ge, Qing-Hai
and Yao-Yang, that found no divergences in the chiral Schwinger
model considered in the gauge invariant formalism.
\end{abstract}


\section{Introduction}

Gauge theories are nowadays responsible for the description of
elementary interactions \cite{Abers-Lee}. One of the main
requirements of the standard model is that of anomaly
cancellation, without which it is not known how to perform
perturbative calculations \cite{t'Hooft}. The phenomenologically
achieved equilibrium between the number of families of quarks and
leptons guarantees this cancellation. However, in practice,
nothing prevents the discovery of new kinds of quarks or leptons,
as higher energies are reached. This could threaten the
theoretical consistency of the model and raise questions about the
correctness of the gauge approach to elementary interactions.

However, some features of gauge theories were discovered in the
80's that could put the questions above under a more comfortable
perspective. It was realized that, at least in 2 dimensional gauge
theories, quantum consistency could be reached even for anomalous
theories \cite{Jackiw-Rajaraman} (the anomaly being understood as
occurring in the gauge symmetry). A mechanism of symmetry
restoration seemed to be operating in the background, becoming
explicit through the natural introduction of a new set of degrees
of freedom, available only at quantum level, the so-called
{\it{Wess-Zumino fields}} \cite{FaddeevS,SchapoV,HaradaT}. Since
then, this mechanism has been intensively studied \cite{Abdalla}
although the achievements have been little in four dimensions
(see, however, \cite{TDKieu}).

In two dimensions, two strategies have been mainly followed
(taking advantage of the fact that, for this number of dimensions,
exactly soluble models are well known). The first, already
mentioned above, consists in studying the dynamics of the theory
that emerges when one considers explicitly the Wess-Zumino fields,
and is called {\it{gauge invariant formalism}} (GIF). The other
\cite{LinharesR}, takes into account explicitly the dynamics of
the longitudinal part of the gauge field, given by the anomaly. It
is called {\it{gauge non-invariant formalism}} (GNIF). In both
formalisms one ends with a gauge invariant theory, whether one
integrates over the fermions and the Wess-Zumino fields (in GIF)
or over the fermions and the longitudinal part of the gauge field
(in GNIF). This is achieved regardless of the regularization
method employed, which can preserve or not gauge invariance in
intermediate computations.

This fact suggests that one should consider gauge theories in
schemes wider than usual. As this mechanism of ``restoration" of
gauge symmetry is acting, there is {\it a priori} no reason to
consider a fixed value for the Jackiw-Rajaraman parameter (that
value which  preserves gauge invariance in intermediate
calculations) which appears precisely as a manifestation of
regularization ambiguities. In fact, for theories involving chiral
fermions, there is no value for this parameter that can preserve
intermediate gauge invariance. However one ends up with an
effective action which is explicitly gauge invariant
\cite{HaradaT}.

Another well-known fact is that fermionic correlation functions
are divergent if one considers a gauge non-invariant scheme for
regularizing the theory \cite{GirottiRR,Boianovsky,Chineses}. A
fermion wave function renormalization is enough to render the
theory finite, but several subtleties appear that make it very
hard to be done exactly, making that the label ``exactly soluble"
be dependent on technical advances \cite{tiao-rodolfo}. The main
problems are: 1) to identify precisely the origin of the
divergences and 2) to learn how to deal with them. The second
problem is treated in \cite{tiao-rodolfo}. This paper addresses
itself to the detailed examination of the origin of the
divergences. We tried to localize this origin in both GIF and
GNIF, as a means to check {\it a posteriori} the self-consistency
of the two approaches and to have a more clear picture of what is
happening.

In particular, we found results that are in explicit contradiction
to the ones found by Jian-Ge, Qing-Hai and Yao-Yang in
\cite{chineses-JP}. In their paper, they found no divergences in
the chiral Schwinger model, when considered in the gauge invariant
formalism. At the conclusions, we briefly comment their paper and
the reasons that conducted them to this mistake.

This paper is organized as follows: in section 2 we review the GIF
and the GNIF, applying the results to study the Schwinger model
(in section 3) and the chiral Schwinger model (in section 4)
perturbatively in both formalisms. In these sections we show that
the divergences in fermionic Green's functions have a
non-perturbative nature, in both models and in both formalisms. We
present our conclusions in section 5.

\section{GIF and GNIF}

The models to be studied are defined by the Lagrangian densities,
\begin{eqnarray} \label{lag} {\cal
L}[\psi,\overline\psi,A]=-\frac{1}{4}F_{\mu\nu}F^{\mu\nu} +
\overline\psi(\,i\partial\!\!\!\slash +e\,A\!\!\!\slash\,P\,)\psi
,
\end{eqnarray} where
\begin{eqnarray} \label{projetor} P=\left\{\begin{array}{l}
{\mathbf 1} \;,\;\mbox{Schwinger model}\\ P_\pm\;,\;\mbox{chiral
Schwinger model}
\end{array}\right.\nonumber
\end{eqnarray} The fermionic determinant is computed exactly in both
cases,
\begin{eqnarray}\label{detf}
\det(i\partial\!\!\!\slash+eA\!\!\!\slash\,P)=\exp\left(iW[A_\mu]
\right)=\int\!\!
d\psi d\overline\psi \exp\left(i\int\!\!
dx\;\overline\psi(i\partial\!\!\!\slash+eA\!\!\!\slash\,P)\psi\right).
\end{eqnarray} In general, the determinant has to be calculated
through a regularization  prescription. For the Schwinger model it
is possible to do it in an gauge invariant way, but this is not
the case for the chiral coupling. We will calculate the
determinant using gauge non-invariant prescriptions, even though,
for the Schwinger model, no authentic gauge anomalies are obtained
in this way \cite{Zinn-Justin}.

To define a free propagator for the field $A_\mu$ from
(\ref{lag}), it is usual to introduce a gauge fixing condition.
However, the situation here is not the same as in usual gauge
theories. As the theory after quantizing fermion fields is gauge
non-invariant, the results of its quantization will depend,
potentially, on the gauge fixing condition used. This prevents us
from using a gauge fixing condition in (\ref{lag}). As we will see
below, this problem can be easily bypassed, both in GIF and in
GNIF.

Let us consider the functional integration over the fermion fields
in (\ref{detf}), performing the following change in fermionic
variables \begin{eqnarray}  \label{psipsi0}
\begin{array}{lcl}
\psi &\rightarrow & \psi=g\psi^g \\ \overline\psi &\rightarrow &
\overline\psi=\overline{{\psi}^g} g^\dag,
\end{array}
\end{eqnarray} where $g$ belongs to the gauge group under which the
fermions transform. In general, we expect that the fermionic
measure would not be invariant under this transformation, unless
we use explicitly a gauge invariant prescription to define it
(which is impossible, in the chiral Schwinger model). Then, we
have in general, \begin{eqnarray} \label{mednicf} d\psi
d\overline\psi =J[A_\mu,g]d\psi^g d\overline{\psi^g} ,
\end{eqnarray} where $J[g, A_\mu]$ is the Jacobian of the
transformation. Having computed the fermion determinant, it is
possible to obtain this Jacobian easily \cite{HaradaT},
\begin{eqnarray}\label{mednicf1}
J[A_\mu,g]&=&e^{i\left(W[A_\mu]-W[A^{g}_\mu]\right)},
\end{eqnarray} where,
\begin{eqnarray} A^{g}_\mu&=&g^{-1}A_\mu g+\frac{i}{e}g^{-1}
\partial_\mu g. \end{eqnarray}
We observe that, if we use a prescription which preserves the
gauge symmetry of $W[A_\mu]$, we obtain $J[g, A_\mu]=1$, according
to our expectations.

These are the basic facts that lie below the formalisms that we
are going to review in the next sections.


\subsection{The gauge invariant formalism}

The generating functional of the theory~(\ref{lag}) is given by
the following definition \begin{eqnarray} \label{gerf1}
Z[\eta,\overline\eta,J]=N\int\!\! dA_\mu d\psi
d\overline\psi\;\exp\left[\;i\int\!\! dx\left(\;{\cal
L}[\psi,\overline\psi,A]+\overline\eta\psi+\overline\psi\eta
+J{\cdotp\!}A \;\right)\right]. \end{eqnarray} We return to the
problem of defining a free propagator for the field $A_\mu$. We
notice that, if the theory were gauge invariant at quantum level,
we should use Faddeev-Popov's technique \cite{Zinn-Justin} to
obtain a well defined functional integration. Harada and Tsutsui
~\cite{HaradaT}, and Babelon, Schaposnik e Viallet~\cite{SchapoV},
observed that this is not necessary (in fact, it is redundant)
when the theory is not gauge invariant at quantum level because,
in this case, different gauge orbits of $A_\mu$ give different
contributions to the effective action. However, Faddeev-Popov's
technique can still be applied, as it consists of multiplication
by $1$, expressed as \begin{eqnarray} \label{f-p}
1=\Delta_f[A_\mu]\int\!\! dg\;\delta(f[A^g_\mu]). \end{eqnarray}
In the above formula, $dg$ represents the invariant measure over
the gauge group $\mathbf G$, $g\in \mathbf G$ and $f[A_\mu]$ is
the gauge fixing condition. Thus, as usual, we insert~(\ref{f-p})
in equation (\ref{gerf1}) and change integration variables in the
bosonic sector $A_\mu \rightarrow A^{g^{-1}}_\mu$ ($dA_\mu$ and
$\Delta_f[A_\mu]$ are gauge invariant by construction). The
generating functional (\ref{gerf1}) becomes \begin{eqnarray}
\label{gerf3} Z[\eta,\overline\eta,J]&=&N\int\!\! dA_\mu d\psi
d\overline\psi dg\;\Delta_f[A_\mu]\,\delta(f[A_\mu]) \\ & &
\times\exp\left[\;i\int\!\! dx\left(\;{\cal L}
[\psi,\overline\psi,A^{g^{-1}}]+ \overline\eta\psi+
\overline\psi\eta +J^\mu A^{g^{-1}}_\mu \;\right)\right].\nonumber
\end{eqnarray} Now, we redefine the fermionic fields according to
the rule
\begin{eqnarray} \label{changfer}
\begin{array}{lcl}
\psi &\rightarrow & \psi=g\psi^g \label{psipsi},\\ \overline\psi
&\rightarrow & \overline\psi= \overline{\psi^g}g^{\dag},
\end{array}
\end{eqnarray} and we see that the Lagrangian  returns to its original
form.
However, as the measure has not necessarily been defined in a
gauge invariant way, it is not invariant under the
transformation~(\ref{changfer}), but changes as we saw in
(\ref{mednicf1}), \begin{eqnarray} d\psi d\overline\psi =d\psi^g d
\overline{\psi^g} e^{i\alpha[A_\mu,g^{-1}]}, \end{eqnarray} where
$\alpha[A_\mu,g^{-1}]=W[A^{g^{-1}}_\mu]-W[A_\mu]$ is the
Wess-Zumino  action.
 Thus, we obtain the following expression for the generating functional,
\begin{eqnarray} \label{gerf4} Z[\eta,\overline\eta,J] & = & N\int\!\!
dA_\mu d\psi d\overline\psi dg\;
\Delta_f[A_\mu]\delta(f[A_\mu])\;\exp\left(i\alpha[A_\mu,g^{-1}]\right)
\\
& &\hspace{-0.5cm}\times\exp\left[\,i\int\!\! dx\left({\cal
L}[\psi,\overline\psi,A]+\overline\eta g\psi+\overline\psi
g^\dag\eta + J{\cdotp\!}A^{g^{-1}}\right)\right].\nonumber
\end{eqnarray} Now we can define a free propagator for the $A_\mu$
field, by exponentiating the $\delta-$function as in the ordinary
situation. The quantization of the theory is, thus, independent of
the choice of the gauge fixing condition, as in the usual
formulation of Faddeev-Popov. We will use the Lorentz gauge fixing
condition, $f[A_\mu]=\frac{1}{\sqrt \xi}\partial\cdotp\!A\;$ and
we will absorb $\Delta_f[A]$ in the normalization constant
(because the theories are Abelian and the Faddeev-Popov's ghost
fields decouple). Doing this, we arrive at \begin{eqnarray}
\label{gfunt} Z[\eta,\overline\eta,J] & = & N' \int\!\! dA_\mu
d\psi d\overline\psi dg\,\exp\left(i\alpha[A_\mu,g^{-1}]\right)\\
& &\times\exp\left[i\int\!\! dx\left({\cal
L}[\psi,\overline\psi,A]-\frac{1}{2\xi}
(\partial{\cdotp}A)^2+\overline\eta g\psi+\overline\psi g^\dag
\eta + J{\cdotp\!}A^{g^{-1}}\right)\right].\nonumber
\end{eqnarray} This equation will be the starting point for the
perturbative analysis of the theory defined by (\ref{lag}). As we
are in a gauge non-invariant theory, the source $J^\mu$ will have
its divergence $\partial_\mu J^\mu \neq 0$, in general. Another
thing that must be indicated is that, in the process of defining
the free propagator of $A_\mu$, the theory acquires an additional
degree of freedom, given by the Wess-Zumino field, which, in the
end, interacts with the fermion fields through the fermionic
sources. This interaction is very complicated, and prevents the
exact calculation of $Z[\eta,\overline\eta,J]$. However, the exact
calculation of an arbitrary correlation function is possible, at
least in principle (once one renormalizes the divergences to be
found in the next sections). The possibility of defining correctly
a free propagator for $A_\mu$ is essential to perform a
perturbative analysis of the theory, and thus, to be able to see
if the ultraviolet divergences that appear in the fermionic
Green's functions have a perturbative origin or not.


\subsection{The gauge non-invariant formalism}

Another approach to the perturbative problem is commonly called
gauge non-invariant formalism~\cite{LinharesR,HaradaN}. In this
context, we use the fact that the classical decoupling of the
longitudinal part of $A_\mu$ (that can be obtained with a gauge
transformation of the fermion fields) does not keep the fermionic
measure invariant, in general. This fact can be exploited to
obtain a perturbative description of the theory, as we will see
below.

Let us separate the field $A_\mu$ in its longitudinal and
transverse parts, as usual, \begin{eqnarray} eA_\mu&=&
\partial_\mu \rho-\tilde\partial_\mu\phi, \end{eqnarray} and substitute
the expression above into the generating functional~(\ref{gerf1})
\begin{eqnarray} \label{ger2} Z[\eta,\overline\eta,J]&\!\!\!=&\!\!\!
N\int\!\! d\rho d\phi d\psi d\overline\psi\;\exp\left(i\int\!\!
dx\;{\cal L}[\psi,\overline\psi,\frac{1}{e} \partial_\mu
\rho-\frac{1}{e}\tilde\partial_\mu\phi] \right) \nonumber \\  &
&\times\exp\left[i\int\!\!
dx\left(\overline\eta\psi+\overline\psi\eta +\frac{1}{e}J_\mu
\partial^\mu\rho-\frac{1}{e}J_\mu
\tilde\partial^\mu\phi\;\right)\right]. \end{eqnarray} We see
that, as the classical action is gauge invariant, the longitudinal
part of the field $A_\mu$ (the field $\rho$) does not have a
kinetical term, apparently appearing as an auxiliary field.
Classically it is possible to remove completely this field from
the sourceless part of the action through the following
transformation
\begin{eqnarray}\label{mudfer}
\begin{array}{lcl}
\psi&\rightarrow&\psi=g\psi'\\
\overline\psi&\rightarrow&\overline\psi=\overline{\psi'} g^\dag.
\end{array}
\end{eqnarray} with $g=e^{i\rho P}$. If the measure were invariant under
this transformation, we would have a linear dependence on $\rho$,
that would render its integration undefined (in the non-anomalous
case, that is why we have to use Faddeev-Popov's method: to
generate a kinetical term for $\rho$). However, the fermionic
measure, as we saw, is not invariant under (\ref{mudfer}), but
changes as
\begin{eqnarray} d\psi d\overline\psi = d\psi' d\overline{\psi'}
e^{i\alpha[\rho,\phi]}\; , \end{eqnarray} where
$\alpha[\rho,\phi]=W[\frac{1}{e}\partial_\mu\rho-\frac{1}{e}
\tilde\partial_\mu\phi]-W[-\frac{1}{e}\tilde\partial_\mu\phi]$.

The generating functional~(\ref{ger2}) then acquires the following
form\begin{eqnarray} \label{ger333}
Z[\eta,\overline\eta,J]&=&N\int\!\! d\rho d\phi d\psi
d\overline\psi\;\exp\left(i\alpha[\rho,\phi]+i \int\!\! dx\; {\cal
L}[\psi,\overline\psi,-\frac{1}{e}\tilde\partial_\mu\phi]\right)\\
& &\times\exp\left[i\int\!\! dx\left(\overline\eta
g\psi+\overline\psi g^\dag \eta +\frac{1}{e}J_\mu\partial^\mu\rho
-\frac{1}{e}J_\mu\tilde\partial^\mu \phi\right)\right].\nonumber
\end{eqnarray} As we are going to see in the next two sections,
the $\alpha$ term contains a kinetical term for the $\rho$ field
which allows us to treat the theory perturbatively. We notice that
the coupling of $\rho$ to the fermion fields is done through the
fermionic sources and is not minimal anymore. However, the theory
in (\ref{ger333}) now admits a perturbative analysis, as we will
explicitly show for the two models mentioned in the beginning.


\section{Schwinger model}

\subsection{Perturbative analysis in GIF}

The Schwinger model is defined by setting $P={\mathbf 1}$ (see
equation (\ref{projetor})). A typical gauge group element can be
parameterized by a field $\theta$ as $g=e^{i\theta}$. The
Wess-Zumino action is
\cite{Dias-Linhares}\begin{eqnarray}\label{wzr}
\alpha(A_\mu,\theta)&=&\frac{(a-1)}{2\pi}\int\!\!
dx\left(\;\frac{1}{2}\partial_\mu \theta \partial^\mu \theta -
e\theta\partial_\mu A^\mu\;\right). \end{eqnarray} Notice that,
for $a=1$, the Wess-Zumino action is zero and we have a Jacobian
equal to one, which characterizes quantum gauge invariance. We are
going to compute bosonic and fermionic Green's functions
perturbatively, looking for the appearance of divergences.
\subsubsection{Photon Propagator}
If we take two functional derivatives with respect to $J_\mu(x)$
and $J_\nu(y)$ of (\ref{gfunt}) we obtain,  \begin{eqnarray}
\label{profot22} \langle
0|TA_\mu(x)A_\nu(y)|0\rangle&\!\!\!\equiv&\!\!\! _\theta\langle
0|TA_\mu(x) A_\nu(y)|0\rangle_\theta+\frac{1}{e} \;_\theta\langle
0|TA_\mu(x)\partial^y_\nu \theta(y)|0\rangle_\theta\;+\\ &
&+\;\frac{1}{e}\;_\theta\langle0|T\partial^x_\mu\theta(x)A_\nu(y)|0
\rangle_\theta +\frac{1}{e^2}\;_\theta\langle
0|T\partial^x_\mu\theta(x)\partial^y_\nu
\theta(y)|0\rangle_\theta\;.\nonumber \end{eqnarray} In the
equation above we see that the photon propagator in the original
theory is expressed as a sum of propagators referred to another
theory given by an action $S_{\theta}[\psi,\overline\psi,
A_\mu,\theta]$ that includes the $\theta$ field,

\vskip 1cm
\begin{eqnarray}
S_{\theta}[\psi,\overline\psi,A_\mu,\theta]&\!\!\!=&\!\!\!\!\int\!\!
dx\left[-\frac{1}{4}F_{\mu\nu}F^{\mu\nu}-\frac{1}{2\xi}(\partial
{\cdotp}A)^2+
\overline\psi(i\partial\!\!\!\slash+eA\!\!\!\slash)\psi\,+\right.\\
& & \left.\hspace{2.5cm}+\frac{(a-1)}{4\pi}\partial_\mu \theta
\partial^\mu \theta - \frac{(a-1)}{2\pi}e\theta\partial_\mu
A^\mu\right]\nonumber \end{eqnarray} We will denote the
expectation values in this theory by
$_\theta\langle\;\;\rangle_\theta$. From
$S_{\theta}[\psi,\overline\psi,A_\mu,\theta]$ we easily obtain
Feynman rules and compute the necessary expectation values. In the
case that we are considering here, it will be possible to add up
the perturbative series and compare with the exact result. In the
following we exhibit the relevant Feynman diagrams and their
result after computation:

a) $A_\mu-$propagator, $\;_\theta\langle
0|TA_\mu(x)A_\nu(y)|0\rangle_\theta$ :
\begin{eqnarray}\label{serieic}
\hspace{-6.24cm}\scalebox{1}{\includegraphics[86,550][300,650]
{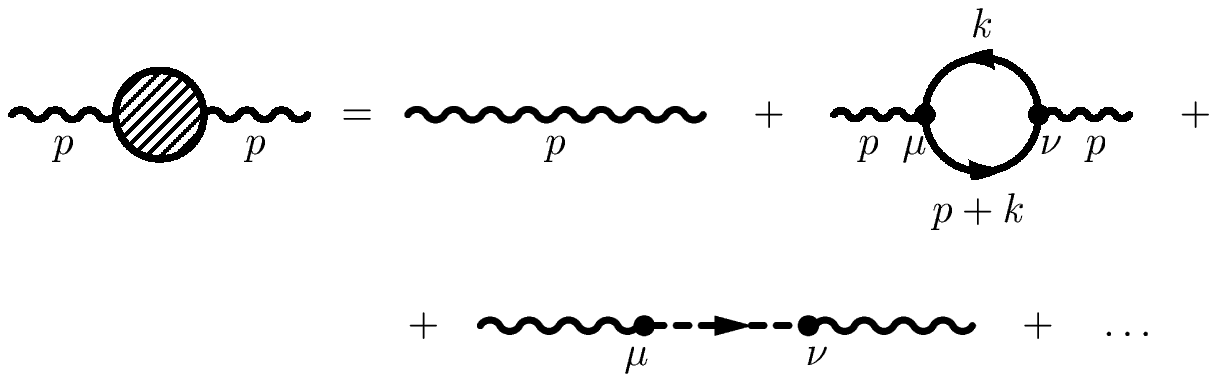}}
\end{eqnarray}
The fermionic loop in~(\ref{serieic}) has to be computed using
some regularization, which may be of the Pauli-Villars kind, for
example (see also \cite{Dias-Linhares}). However, once this
regularization is employed, as is well known, the result is
finite. Adding the contributions up to the third graphic, we get a
result valid at order $e^2$
\begin{eqnarray} -\frac{i\,e^2(a+1)}{2\pi
p^4}\left(g_{\mu\nu}-\frac{p_\mu p_\nu}{p^2}\right).\nonumber
\end{eqnarray} Noticing that (\ref{serieic}) is a geometrical
series with $n-$th term ($n\geqslant 1$) \begin{eqnarray}
\frac{-i}{p^2}\left(\frac{e^2(a+1)}{2\pi p^2}\right)^n
\left(g_{\mu\nu}- \frac{p_\mu p_\nu}{p^2}\right),\nonumber
\end{eqnarray} we can easily add the series~(\ref{serieic}) to
obtain the complete $A_\mu-$propagator in momentum space,
\begin{eqnarray} _\theta\langle
0|TA_\mu(p)A_\nu(-p)|0\rangle_\theta&=&\frac{-i}{p^2-
\frac{e^2(a+1)}{2\pi}}\left(\,g_{\mu\nu}-\frac{p_\mu
p_\nu}{p^2}\,\right) -\frac{i\xi p_\mu p_\nu}{p^4} \end{eqnarray}

b) $\theta-$propagator, $\;_\theta\langle
0|T\theta(x)\theta(y)|0\rangle_\theta$ : \begin{eqnarray}
\hspace{-6.24cm}\scalebox{1}{\includegraphics[86,618][300,650]
{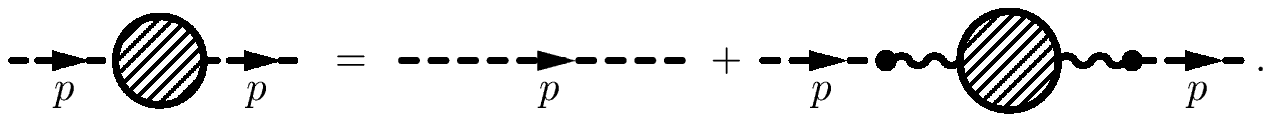}}\nonumber \end{eqnarray} Computing these two
contributions, we get
\begin{eqnarray} _\theta\langle
0|T\theta(p)\theta(-p)|0\rangle_\theta &=&
i\left(\frac{2\pi}{a-1}\frac{1}{p^2}-\frac{\xi e^2}{p^4}\right).
\end{eqnarray}

c) Mixing terms, $\;_\theta\langle
0|T\theta(x)A_\mu(y)|0\rangle_\theta$ and $_\theta\langle
0|TA_\mu(x)\theta(y)|0\rangle_\theta$ :

The two terms that contribute are: \begin{eqnarray}
\hspace{-6.24cm}\scalebox{1}{\includegraphics[86,618][300,650]
{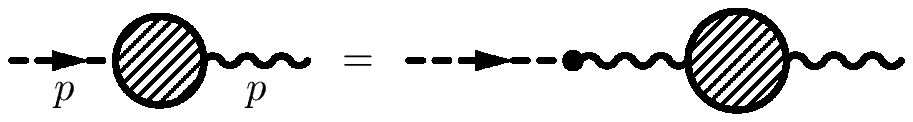}}\nonumber
\end{eqnarray} \begin{eqnarray} _\theta\langle
0|T\theta(p)A_\mu(-p)|0\rangle_\theta&=&-\frac{\xi e p_\mu}{p^4}
\end{eqnarray} \begin{eqnarray}
\hspace{-6.24cm}\scalebox{1}{\includegraphics[86,618][300,650]
{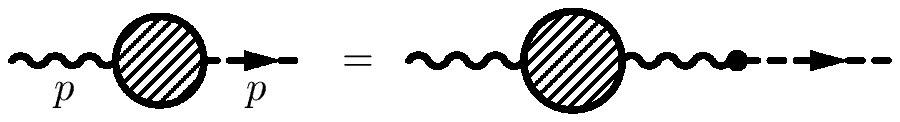}}\nonumber
\end{eqnarray} \begin{eqnarray} _\theta\langle
0|TA_\mu(p)\theta(-p)|0\rangle_\theta&=&\frac{\xi e p_\mu}{p^4}
\end{eqnarray} Adding all results as in (\ref{profot22}), we obtain
the full photon propagator for the theory,
\begin{eqnarray}\label{propfotex}
i\langle0|TA_\mu(p)A_\nu(-p)|0\rangle=\frac{1}{p^2-
\frac{e^2}{2\pi}(a+1)}
\left(g_{\mu\nu}-\frac{p_\mu
p_\nu}{p^2}\right)-\frac{2\pi}{e^2(a-1)} \frac{p_\mu p_\nu}{p^2}.
\end{eqnarray} This result agrees exactly with which is obtained by
non-perturbative methods \cite{Mitra-Rahaman} (taking into account
that the $\overline{a}$ parameter there is related to ours as
$\overline{a}=(a-1)/2$).

\subsubsection{Fermion propagator}

From~(\ref{gfunt}), we can take functional derivatives with
respect to the fermionic sources $\overline\eta(x)$ e $\eta(y)$
and compute the fermion propagator as \begin{eqnarray} \langle
0|T\psi(x)\overline\psi(y)|0\rangle &=& N'\!\int\!\! dA_\mu
d\theta d\psi d\overline\psi \;\;\psi(x)\overline\psi(y)\nonumber
\\  &
&\times\exp\left(i\,S_{\theta}[\psi,\overline\psi,A_\mu,\theta]+
\!\int\!\! dz\;\theta(z)j(z,x,y)\right) \end{eqnarray} where
$j(z,x,y)=\delta(z-x)-\delta(z-y)$. Integrating over the $\theta$
field, we obtain \begin{eqnarray} \label{petfer22} \langle
0|T\psi(x)\overline\psi(y)|0\rangle =
\exp\left\{-\frac{2\pi\,i}{a-1} \int\!\!
\frac{dk}{(2\pi)^2}\;\frac{1-e^{-ik{\cdotp}(x-y)}}{k^2}\right\}\;
G_p(x-y),
\end{eqnarray} where the exponential contains a divergence already
found elsewhere \cite{tiaomarcelo,Mitra-Rahaman} which is
generated by the integration of the field $\theta$. This
divergence is not cancelled by the normalization factor $N'$, as
it is induced by the presence of $j(z,x,y)$ (which, in turn, is
generated by the functional derivations, absent in the
normalization factor). $G_p(x-y)$ is defined from the remaining
functional integration in terms of the fields $A_\mu$, $\psi$ and
$\overline\psi$,
\begin{eqnarray}\label{petfer2} G_p(x-y)&=&N''\int\!\! dA_\mu d\psi
d\overline\psi
\;\;\psi(x) \overline\psi(y)\\ & & \times\exp\left\{i\int\!\!
dz\left(\;\frac{1}{2}A_\mu H^{\mu\nu}_\xi
A_\nu+\overline\psi(i\partial\!\!\!\slash
+eA\!\!\!\slash)\psi+eA_\mu l^\mu(z,x,y)\right)\right\}\nonumber
\end{eqnarray} where
\begin{eqnarray}
l^\mu(z,x,y)=\partial^\mu_z[\,D_F(z-x)-D_F(z-y)\,], \nonumber
\end{eqnarray} and
\begin{eqnarray} \label{HHH}
H^{\mu\nu}_\xi&=&g^{\mu\nu}\square+(\frac{1}{\xi}-1)\partial^\mu
\partial^\nu
-\frac{e^2}{2\pi}(a-1)\frac{\partial^\mu\partial^\nu}{\square}.
\end{eqnarray} From (\ref{HHH}), we obtain an effective free propagator
for $A_\mu$, that we call $h^\xi_{\mu\nu}$ \begin{eqnarray}
h^\xi_{\mu\nu}(k)=-\frac{i}{k^2}\left(g_{\mu\nu}-\frac{k_\mu
k_\nu}{k^2}\right)-\frac{i\xi k_\mu k_\nu }{k^2[k^2+\frac{\xi
e^2}{2\pi}(a-1)]}.\nonumber \end{eqnarray} Its ultraviolet
behavior is of the form $k^{-2}$. The Feynman rules to calculate
the fermion self-energy are now given by the action appearing in
the functional integration (\ref{petfer2}). It is now easy to
calculate $G_p(x-y)$ to any desired loop order. We limit ourselves
to the 1-loop contribution to the fermion self-energy,
\begin{eqnarray} \label{selffer1}
-i\Sigma_p(p)&=&i\,p\!\!\!\slash\left[\frac{e^2}{4\pi
p^2}-\frac{1}{2(a-1)} \ln\left(1+\frac{\xi e^2}{2\pi
p^2}(a-1)\right)\right],
\end{eqnarray} that is finite, as well as all the other diagrams
that enter in the computation of $G_p(x-y)$. So, the only source
of divergences in (\ref{petfer22}) is the integration over the
$\theta$ field, which is done exactly, outside the perturbative
level.

A little further reflection shows quickly that the same is true
for all fermionic Green's functions: they all exhibit a
divergence, originating in the integration over the Wess-Zumino
field, being finite {\it modulo} this problem.


\subsection{Perturbative analysis in GNIF}

Now we start from (\ref{ger333}), where \begin{eqnarray} {\cal
L}[\psi,\overline\psi,-\frac{1}{e}\tilde\partial_\mu\phi]&=&
\frac{1}{2e^2}\phi\,\square^2\phi+\overline\psi(i\partial\!\!\!\slash-
\tilde\partial\!\!\!\slash\phi)\psi, \end{eqnarray} having
$g=e^{i\rho}$ and a Jacobian given by \begin{eqnarray}
\label{aaapha}
\alpha(\rho,\phi)=\exp\left(\frac{i\,(a-1)}{4\pi}\int\!\!
dx\;\partial_\mu\rho
\partial^\mu\rho\right).
\end{eqnarray}
\subsubsection{Photon propagator}
The propagator for the $A_\mu$ field has to be expressed in terms
of propagators for the $\rho$ and $\phi$ fields. In (\ref{ger333})
we take two functional derivatives with respect to $J_\mu$ and
$J_\nu$, put all the sources to zero and we are left with
\begin{eqnarray} \label{pertfoton11} \langle
0|TA_\mu(x)A_\nu(y)|0\rangle&\!\!\!=&\!\!\!\frac{1}{e^2}
\left(_l\langle
0|T\partial^x_\mu\rho(x)\partial^y_\nu\rho(y)|0\rangle_l\,+
\,_l\langle0|T\tilde\partial^x_\mu\phi(x)\tilde\partial^y_\nu
\phi(y) |0 \rangle_l\right).\nonumber \\  \end{eqnarray}
$_l\langle\,|\,\rangle_l$ refers to expectation values calculated
using the effective action
\begin{eqnarray}\label{sefef} S_{l}[\psi,\overline\psi,\rho,\phi]
&=&\!\!\!\int\!\!
dx\left(\frac{1}{2e^2} \phi \square^2 \phi +
\frac{(a-1)}{4\pi}\partial_\mu\rho
\partial^\mu\rho+\overline\psi(i \partial\!\!\!\slash -
\tilde\partial\!\!\!\slash\phi
)\psi\right). \end{eqnarray} The photon propagator is split into a
sum of two propagators of the fields $\rho$ and $\phi$, whose
dynamics is described by the action above. We see in
$S_{l}[\psi,\overline\psi,\rho,\phi]$ that $\rho$ is a free field,
which implies that the mixed propagators $\;_l\langle
0|T\partial^x_\mu\rho(x)\tilde\partial^y_\nu\phi(y)|0\rangle_l\;$
and $_l\langle
0|T\tilde\partial^x_\mu\phi(x)\partial^y_\nu\rho(y)|0\rangle_l\;$
are null. With the Feynman rules generated from
$S_{l}[\psi,\overline\psi,\rho,\phi]$, we can calculate the photon
propagator for the theory in a perturbative way. We show these
results below:

a) $\phi-$propagator, $\;_l\langle 0|T\phi(x)\phi(y)|0\rangle_l$:
\begin{eqnarray}\label{series2}
\hspace{-6.24cm}\scalebox{1}{\includegraphics[86,545][300,650]
{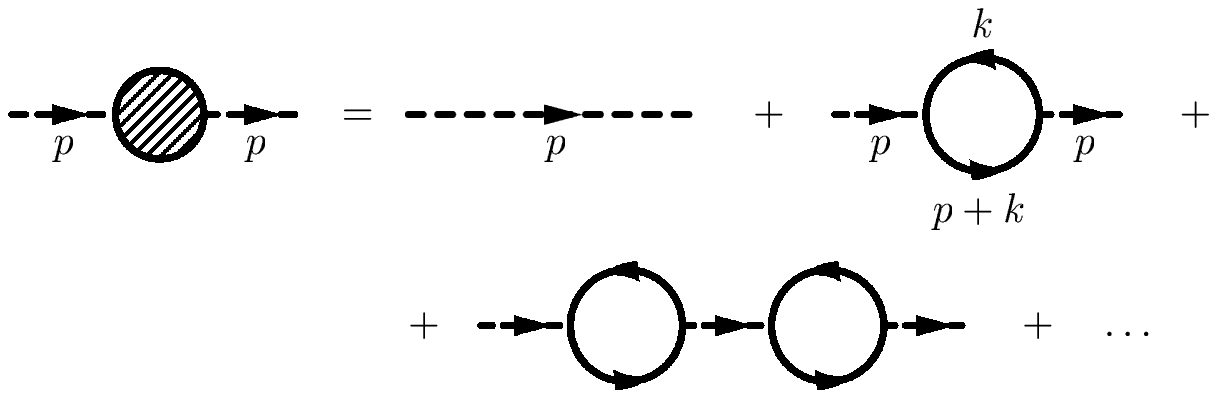}}
\end{eqnarray}
The fermionic loop (\ref{series2}) is calculated with a
Pauli-Villars regularization prescription, as before. Its
contribution, in this case, is \begin{eqnarray}
-\frac{i(a+1)}{2\pi}p^2.\nonumber \end{eqnarray} Adding the series
(\ref{series2}), we get the propagator of the $\phi$ field
\begin{eqnarray} _l\langle
0|T\phi(p)\phi(-p)|0\rangle_l=\frac{ie^2}{p^2(p^2-
\frac{e^2(a+1)}{2\pi})} \end{eqnarray}

b) $\rho-$propagator, $\;_l\langle 0|T\rho(x)\rho(y)|0\rangle_l$:
as the $\rho$ field in (\ref{sefef}) is a free field, its
propagator is calculated directly, \begin{eqnarray} _l\langle
0|T\rho(p)\rho(-p)|0\rangle_l&=&\frac{2\pi i}{(a-1)\;p^2}.
\end{eqnarray}

The photon propagator is then obtained from equation
(\ref{pertfoton11}),  \begin{eqnarray} i\langle
0|TA_\mu(p)A_\nu(-p)|0\rangle=\frac{1}{k^2-
\frac{e^2(a+1)}{2\pi}}\left(g_{\mu\nu}-\frac{k_\mu
k_\nu}{k^2}\right)-\frac{2\pi}{e^2(a-1)}\frac{k_\mu k_\nu}{k^2}.
\end{eqnarray} It coincides with the propagator computed by non
perturbative calculation, and with the one calculated previously
(\ref{propfotex}) in the gauge invariant formalism.

\subsubsection{Fermion propagator}

From (\ref{ger333}), we take functional derivatives with respect
to the fermion sources and we obtain, \begin{eqnarray}
\label{pertfermion} \langle
0|T\psi(x)\overline\psi(y)|0\rangle&=&N\int\!\! d\rho d\phi d\psi
d\overline\psi\;\;\psi(x)\overline\psi(y)\\ & &
\times\exp\left(iS_{l}[\psi,\overline\psi,\rho,\phi]+i\int\!\!
dz\; \rho(z) j(z,x,y)\right).\nonumber \end{eqnarray} The term
involving $j(z,x,y)=\delta(z-x)-\delta(z-y)$ is generated when we
perform the above mentioned functional derivatives. The
integration over the $\rho$ field factorizes, but the presence of
$j(z,x,y)$ prevents the absorption of this integration in the
normalization factor $N$. Then (\ref{pertfermion}) becomes, after
$\rho$ integration, \begin{eqnarray} \langle
0|T\psi(x)\overline\psi(y)|0\rangle&=&\exp\left\{-\frac{2\pi\,i}{a-1}
\int\!\!
\frac{dk}{(2\pi)^2}\;\frac{1-e^{-ik{\cdotp}(x-y)}}{k^2}\right\}\;
G_p(x-y)\,.
\end{eqnarray} We observe the presence of an ultraviolet divergence in
the exponential, that do not have perturbative origin (as it comes
from the $\rho$ integration) and coincides with the one calculated
previously in the gauge invariant formalism (see equation
(\ref{petfer22})). The remaining functional integration in
$G_p(x-y)$, given by  \begin{eqnarray} \label{GPPP}
G_p(x-y)&\!\!\!=&\!\!\! N'\!\int\!\! d\phi d\psi
d\overline\psi\;\psi(x)\overline\psi(y)\;\exp\left[i\int\!\!
dx\left(\frac{1}{2e^2} \phi \square^2 \phi +\overline\psi(i
\partial\!\!\!\slash - \tilde\partial\!\!\!\slash\phi )\psi\right)
\right],\nonumber \\  \end{eqnarray} is
finite, that is, the Feynman diagrams generated from it do not
show ultraviolet divergences. From (\ref{GPPP}), the 1-loop
contribution to the fermion self-energy is given by
\begin{eqnarray} \label{rficfnic2} -i\Sigma_p(p)=\frac{ie^2
p\!\!\!\slash}{4\pi p^2}. \end{eqnarray} Since it is finite, the
contributions of higher loop order to the self-energy are also
finite.


\section{Chiral Schwinger model}

\subsection{Perturbative analysis in GIF}

We will perform the same analysis for the chiral Schwinger model,
defined by $P=P_\pm$ (see, again, equation (\ref{projetor})). As
before, $g=e^{i\theta}$, but now the Wess-Zumino action is given
by \cite{Dias-Linhares} \begin{eqnarray}\label{aaqq}
\alpha(A,\theta)=\int\!\! dx\left\{\frac{(a-1)}{8\pi}\partial_\mu
\theta \partial^\mu \theta
-\frac{e\theta}{4\pi}\left[\,(a-1)\partial^\mu A_\mu
-\tilde\partial^\mu A_\mu\,\right] \right\} \end{eqnarray} As
opposed to the case of the Schwinger model, there is no value of
$a$ which can turn this action to zero, and this is the
distinctive sign of the gauge anomaly. Again, we will compute
Green's functions perturbatively, observing similarities and
differences in comparison to the vectorial coupling case.

\subsubsection{Photon propagator}

From (\ref{lag}), (\ref{gfunt}) and (\ref{aaqq}), we obtain the
following expression for the photon propagator of the theory
\begin{eqnarray} \label{fotqui} \langle 0|TA_\mu(x)A_\nu(y)|0
\rangle &\equiv&{_\theta\langle} 0|TA_\mu(x)A_\nu(y)|0
\rangle_\theta+\frac{1}{e}{_\theta\langle}
0|TA_\mu(x)\partial^y_\nu\theta(y)|0 \rangle_\theta+\\ & &
+\frac{1}{e}{_\theta\langle} 0|T\partial^x_\mu\theta(x)A_\nu(y)|0
\rangle_\theta+ \frac{1}{e^2}{_\theta\langle}
0|T\partial^x_\mu\theta(x)\partial^y_\nu\theta(y)|0\rangle_\theta
\nonumber \end{eqnarray} The propagator of the original photon is
again a sum of propagators which are referred to an effective
theory $S_{\theta}[\psi,\overline\psi, A_\mu,\theta]$  that
includes the $\theta$ field. We denote the expectation values in
this theory by $_\theta\langle\,|\,\rangle_\theta$. This effective
action $S_{\theta}[\psi,\overline\psi,A_\mu,\theta]$ is given by
\begin{eqnarray}
S_{\theta}[\psi,\overline\psi,A_\mu,\theta]&\!\!\!=&\!\!\!\!\int\!\!
dx\left[-\frac{1}{4}F_{\mu\nu}F^{\mu\nu}-\frac{1}{2\xi}
(\partial{\cdotp}A)^2+
\overline\psi(i\partial\!\!\!\slash+eA\!\!\!\slash\,P_+)\psi\,+\right.
\\
 & & \left.+\frac{(a-1)}{8\pi}\partial_\mu \theta \partial^\mu
\theta - \frac{(a-1)}{4\pi}e\theta\partial_\mu A^\mu+
\frac{1}{4\pi}e\theta\tilde\partial_\mu A^\mu\right]\nonumber
\end{eqnarray} With Feynman rules obtained from
$S_{\theta}[\psi,\overline\psi,A_\mu,\theta]$, we will show, in
what follows, the perturbative calculation of the propagators that
appear in (\ref{fotqui}).

a) $A_\mu-$propagator \begin{eqnarray}\label{serieq}
\hspace{-6.24cm}\scalebox{1}{\includegraphics[86,550][300,650]
{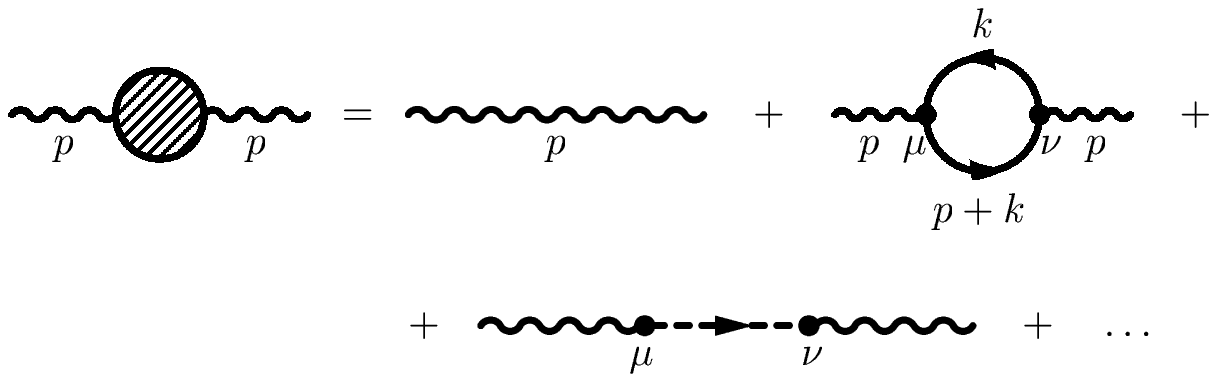}}
\end{eqnarray} The fermionic loop in (\ref{serieq}) is given by
\begin{eqnarray}
-i\Pi_{\mu\nu}(p)=\frac{ie^2}{4\pi}\left[(a+1)g_{\mu\nu}-\frac{2
p_\mu p_\nu}{p^2}-\frac{p_\mu\tilde p_\nu+\tilde p_\mu
p_\nu}{p^2}\right],\nonumber \end{eqnarray} while the third
graphic in (\ref{serieq}) contributes as \begin{eqnarray}
\frac{ie^2}{4\pi}\left[\frac{g_{\mu\nu}}{a-1}-\frac{1+(a-1)^2}{a-1}
\frac{p_\mu p_\nu}{p^2}+\frac{p_\mu\tilde p_\nu+\tilde p_\mu
p_\nu}{p^2}\right].\nonumber \end{eqnarray} Adding both
contributions we obtain, to order $e^2$ \begin{eqnarray}
\frac{i\,e^2\;a^2}{4\pi\,(a-1) }\left[g_{\mu\nu}-\frac{p_\mu
p_\nu}{p^2}\right]. \end{eqnarray} It is easy to see that
(\ref{serieq}) is a geometrical series with an order $n$ term
($n\geqslant 1$)
\begin{eqnarray}
\frac{-i}{p^2}\left(\frac{e^2\;a^2}{4\pi\,(a-1)p^2}\right)^n
\left(\;g_{\mu\nu}-\frac{p_\mu p_\nu}{p^2}\;\right).\nonumber
\end{eqnarray} Adding this series (\ref{serieq}), we get
\begin{eqnarray} _\theta\langle
0|TA_\mu(p)A_\nu(-p)|0\rangle_\theta=\frac{-i}{p^2-
\frac{e^2\;a^2}{4\pi\,(a-1)
}} \left(\;g_{\mu\nu}-\frac{p_\mu p_\nu}{p^2}\;\right)-\frac{i\xi
p_\mu p_\nu}{p^4}. \end{eqnarray}

b) $\theta-$propagator \begin{eqnarray}
\hspace{-6.24cm}\scalebox{1}{\includegraphics[86,618][300,650]
{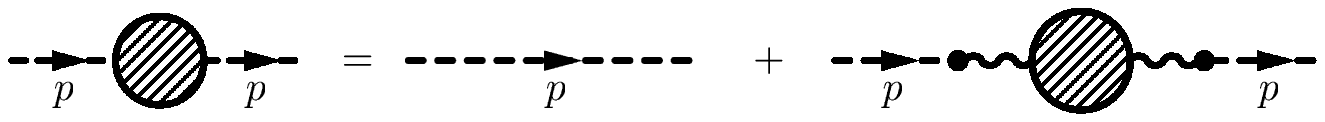}}\nonumber
\end{eqnarray} \begin{eqnarray} _\theta\langle
0|T\theta(p)\theta(-p)|0\rangle_\theta&=&\frac{4\pi}{a-1}
\frac{i}{p^2}-\frac{i\xi
e^2}{p^4}+\frac{ie^2}{p^2(p^2-\frac{e^2\;a^2}{4\pi\,(a-1)})}
\end{eqnarray}

c) Mixed terms $A_\mu-\theta$ \begin{eqnarray}
\hspace{-6.24cm}\scalebox{1}{\includegraphics[86,618][300,650]
{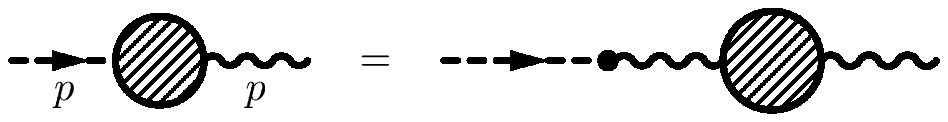}}\nonumber
\end{eqnarray} \begin{eqnarray} _\theta\langle
0|T\theta(p)A_\nu(-p)|0\rangle_\theta&=& -\frac{\xi e
p_\nu}{p^4}+\frac{e}{(a-1)}\;\frac{\tilde
p_\nu}{p^2(p^2-\frac{e^2\;a^2}{4\pi\,(a-1)})} \end{eqnarray}

\begin{eqnarray}
\hspace{-6.24cm}\scalebox{1}{\includegraphics[86,618][300,650]
{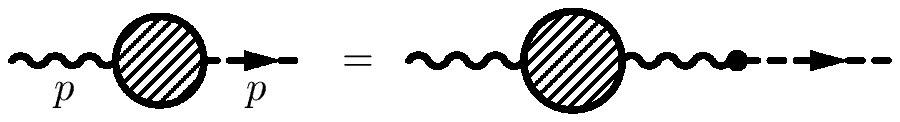}}\nonumber
\end{eqnarray} \begin{eqnarray} _\theta\langle
0|TA_\mu(p)\theta(-p)|0\rangle_\theta&=&\frac{\xi e p_\mu}{p^4}-
\frac{e}{(a-1)}\;\frac{\tilde
p_\mu}{p^2(p^2-\frac{e^2\;a^2}{4\pi\,(a-1)})} \end{eqnarray}

Adding all contributions in (\ref{fotqui}), we obtain the photon
propagator of the theory \begin{eqnarray} \label{fotonexato1q}
i\langle 0|TA_\mu(k)A_\nu(-k)|0 \rangle &=&\\ &
&\hspace{-2.5cm}=\;\frac{1}{k^2-\frac{e^2\;a^2}{4\pi\,(a-1) }}
\left[g_{\mu\nu}-\frac{k_\mu
k_\nu}{a-1}\left(\frac{4\pi}{e^2}-\frac{2}{k^2} \right)
+\frac{k_\mu\tilde k_\nu+\tilde k_\mu
k_\nu}{(a-1)\;k^2}\right]\nonumber
\end{eqnarray}

This is equal to the well known results in the literature
\cite{Jackiw-Rajaraman}. Again we see that, apart from the
regularization of the fermionic loop, there are no perturbatively
induced divergences in this propagator. As in the vectorial case,
only this regularization is enough to furnish a finite result for
the photon propagator.

\subsubsection{Fermion propagator}
From (\ref{lag}), (\ref{gfunt}) and (\ref{aaqq}), we arrive at the
following expression for the fermion propagator \begin{eqnarray}
\langle 0|T\psi(x)\overline\psi(y)|0\rangle=P_-G_F(x-y)+\langle
0|T\psi_+(x)\overline\psi_+(y)|0\rangle, \end{eqnarray} with
$\psi_+=P_+\psi$. The left-handed fermion propagates freely, but
the right-handed one interacts with the vector field $A_\mu$ as
\begin{eqnarray} \label{profermq} \langle
0|T\psi_+(x)\overline\psi_+(y)|0\rangle &=&N\int\!\! dA_\mu d\psi
d\overline\psi d\theta \; \psi_+(x)\overline\psi_+(y)\\ &
&\times\exp\left(i\,S_{\theta}[\psi,\overline\psi,A_\mu,\theta]+
\!\int\!\! dz\;\theta(z)j(z,x,y)\right),\nonumber \end{eqnarray}
with $S_{\theta}[\psi,\overline\psi,A_\mu,\theta]$ given in the
previous section and $j(z,x,y)=\delta(z-x)-\delta(z-y)$.
Integrating over the $\theta$ field, we are left with
\begin{eqnarray}\label{divqic} \langle
0|T\psi_+(x)\overline\psi_+(y)|0\rangle=\exp\left\{-\frac{4\pi\,i}{a-1}
\int\!\!
\frac{dk}{(2\pi)^2}\;\frac{1-e^{-ik{\cdotp}(x-y)}}{k^2}\right\}\;
G^{\prime}_+(x-y).\quad \end{eqnarray} We observe a logarithmic
ultraviolet divergence in the exponential, as in the case of the
anomalous Schwinger model. The remaining functional integration
$G^{\prime}_+(x-y)$ is finite, \begin{eqnarray}\label{fqpfer}
G^{\prime}_+(x-y)&=&N'\int\!\! dA_\mu d\psi d\overline\psi \;
\psi_+(x)\overline\psi_+(y)\\ &
&\hspace{-1cm}\times\exp\left[i\int\!\! dz \left(\frac{1}{2}A_\mu
H^{\mu\nu}
A_\nu+\overline\psi(i\partial\!\!\!\slash+eA\!\!\!\slash\,P_+)
\psi+eA_\mu
l^\mu(z,x,y)\right)\right],\nonumber \end{eqnarray} where
\begin{eqnarray}
H^{\mu\nu}&=&g^{\mu\nu}\left(\square+\frac{e^2}{4\pi(a-1)}\right)+
\left(\frac{1}{\xi}-1\right)\partial^\mu\partial^\nu+\\ & &
-\frac{e^2[(a-1)^2+1]}{4\pi(a-1)}
\frac{\partial^\mu\partial^\nu}{\square}
+\frac{e^2}{4\pi}\frac{\partial^\mu\tilde\partial^\nu+
\tilde\partial^\mu\partial^\nu}{\square},\nonumber \end{eqnarray}
and
\begin{eqnarray} l^\mu(z,x,y)&=&\left(\partial^\mu_z
-\frac{\tilde\partial^\mu_z}{a-1} \right)[D_F(z-x)-D_F(z-y)].
\end{eqnarray} The $A_\mu-$propagator, which enters in
(\ref{fqpfer}), has an ultraviolet behavior as $k^{-2}$. Then, the
1-loop contribution to the fermion self-energy is finite. This
persists to all loop orders.


\subsection{Perturbative analysis in GNIF}
Here we start from (\ref{ger333}), where
\begin{eqnarray}\label{lagmsqni} {\cal
L}[\psi,\overline\psi,-\frac{1}{e}\tilde\partial_\mu\phi]&=&
\frac{1}{2e^2}\phi\,\square^2\phi+\overline\psi(i\partial\!\!\!\slash-
\tilde\partial\!\!\!\slash\phi\,P_+)\psi. \end{eqnarray} Putting
$g=e^{i\rho\,P_+}$, we obtain the $\alpha$ term from the Jacobian
of this gauge transformation \begin{eqnarray} \label{aaaphaq}
\alpha(\rho,\phi)=\exp\left[i\int\!\! dx
\left(\frac{(a-1)}{8\pi}\partial_\mu\rho
\partial^\mu\rho-\frac{1}{4\pi}\partial_\mu\rho\partial^\mu\phi\right)
\right]. \end{eqnarray}

\subsubsection{Photon Propagator}
From equation (\ref{ger333}), considering (\ref{lagmsqni}) and
(\ref{aaaphaq}), we get the photon propagator
\begin{eqnarray}\label{photproq} \langle
0|TA_\mu(x)A_\nu(y)|0\rangle&\!\!\!=&\!\!\!
\frac{1}{e^2}\left({_l\langle}0|T\partial^x_\mu\rho(x)
\partial^y_\nu\rho(y)|0{\rangle_l}
- {_l\langle}
0|T\partial^x_\mu\rho(x)\tilde\partial^y_\nu\phi(y)|0{\rangle_l}+
\right.\nonumber
\\ & &\!\!\!\!\!\!\left.-\, {_l\langle}
0|T\tilde\partial^x_\mu\phi(x)\partial^y_\nu\rho(y) |0{\rangle_l}+
{_l\langle} 0|T\tilde\partial^x_\mu\phi(x)\tilde\partial^y_\nu
\phi(y)|0{\rangle_l}\right). \end{eqnarray} The dynamics is
governed by the effective action
$S_{l}[\psi,\overline\psi,\rho,\phi]$, given by
\begin{eqnarray}\label{sefefq}
S_{l}[\psi,\overline\psi,\rho,\phi]&\!\!\!=&\!\!\!\!\!\!\int\!\!
dx\!\left(\frac{1}{2e^2} \phi \square^2 \phi +\overline\psi(i
\partial\!\!\!\slash - \tilde\partial\!\!\!\slash\phi )\psi+
\frac{(a-1)}{8\pi}\partial_\mu\rho
\partial^\mu\rho- \frac{1}{4\pi}\partial_\mu\rho
\partial^\mu\phi\right).\nonumber \\
& & \end{eqnarray} and $_l\langle\,|\,\rangle_l$ refers to
expectation values calculated in this dynamics. We proceed to the
perturbative calculation of the relevant graphs.

a) $\phi-$propagator \begin{eqnarray}\label{phiprop}
\hspace{-6.24cm}\scalebox{1}{\includegraphics[86,550][300,650]
{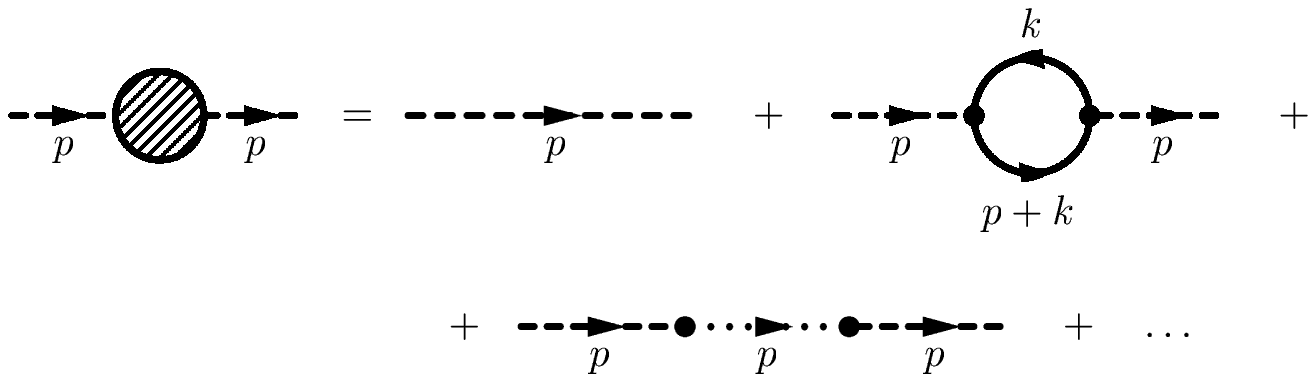}}
\end{eqnarray} The fermionic loop gives \begin{eqnarray}
-\frac{ip^2}{4\pi}(a+1),\nonumber \end{eqnarray}
 and the third graphic contribution is
\begin{eqnarray}
 -\frac{ip^2}{4\pi(a-1)}. \nonumber
\end{eqnarray} The $\phi-$self-energy is \begin{eqnarray}
-i\Sigma_\phi(p)=-\frac{i\,a^2\,p^2}{4\pi\,(a-1)}. \end{eqnarray}
Now, adding the series (\ref{phiprop}), we obtain \begin{eqnarray}
_l\langle 0|T\phi(p)\phi(-p)|0\rangle_l&=&\frac{i\,e^2}{p^2(p^2+
\frac{e^2\;a^2}{4\pi\,(a-1)})} \end{eqnarray}

b) $\rho-$propagator \begin{eqnarray}
\hspace{-6.24cm}\scalebox{1}{\includegraphics[86,635][300,650]
{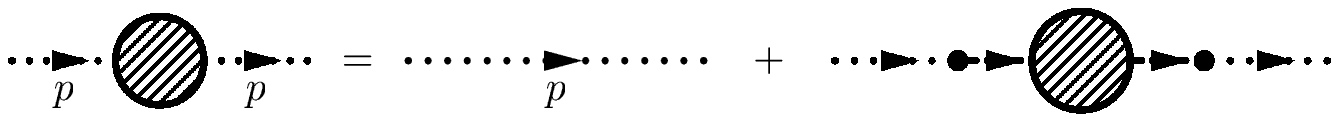}}\nonumber
\end{eqnarray}

\begin{eqnarray} _l\langle
0|T\rho(x)\rho(y)|0\rangle_l&=&\frac{i\,4\pi}{(a-1)p^2}
+\frac{1}{(a-1)^2}\,\frac{i\,e^2}{p^2(p^2-
\frac{e^2\;a^2}{4\pi\,(a-1)})}\quad
\end{eqnarray}

c) Mixed terms $\rho-\phi$ \begin{eqnarray}
\hspace{-6.24cm}\scalebox{1}{\includegraphics[86,540][300,650]
{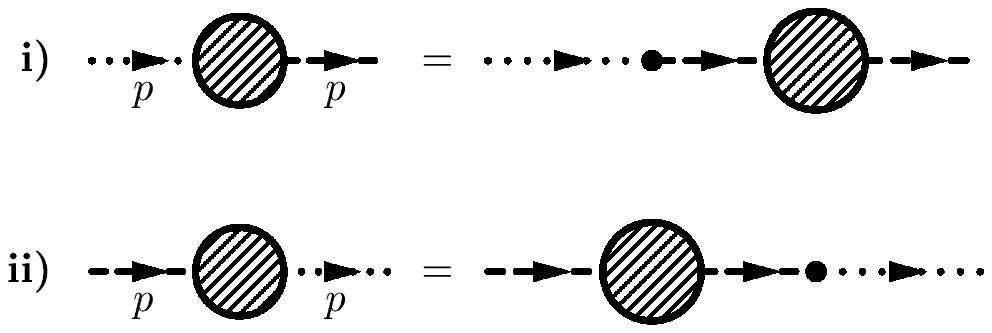}}\nonumber
\end{eqnarray} \begin{eqnarray} \left.\begin{array}{l} {_l\langle}
0|T\rho(p)\phi(-p)|0{\rangle_l}\\ \\ {_l\langle}
0|T\phi(p)\rho(-p)|0{\rangle_l}\end{array} \right\}=
\frac{1}{a-1}\frac{i\,e^2}{p^2(p^2-\frac{e^2\;a^2}{4\pi\,(a-1)})}
\end{eqnarray}
\[ \]
Adding all the contributions we obtain \begin{eqnarray}
\;\;\langle 0|TA_\mu(k)A_\nu(-k)|0\rangle&=&\\ & &\hspace{-2cm}
=\;\frac{-i}{k^2-\frac{e^2\;a^2}{4\pi\,(a-1)}}\left[g_{\mu\nu}-
\frac{k_\mu
k_\nu}{a-1}\left(\frac{4\pi}{e^2}-\frac{2}{k^2}\right)+
\frac{k_\mu\tilde k_\nu+\tilde k_\mu
k_\nu}{(a-1)\;k^2}\right]\nonumber
\end{eqnarray}


\subsubsection{Fermion propagator}
From (\ref{ger333}), we obtain the following expression for the
fermionic propagator \begin{eqnarray} \label{profermq1} \langle
0|T\psi(x)\overline\psi(y)|0\rangle &=& i\,P_-G_F(x-y)+\langle
0|T\psi_+(x)\overline\psi_+(y)|0\rangle \end{eqnarray} where
\begin{eqnarray} \label{pertfermionq} \langle
0|T\psi_+(x)\overline\psi_+(y)|0\rangle&=&N\int\!\! d\rho d\phi
d\psi d\overline\psi\;\;\psi_+(x)\overline\psi_+(y)\\ & &
\times\exp\left(iS_{l}[\psi,\overline\psi,\rho,\phi]+i\int\!\!
dz\; \rho(z) j(z,x,y)\right).\nonumber \end{eqnarray} After
integration over the $\rho$-field, we find the same logarithmic
ultraviolet divergence already found in (\ref{divqic}),
\begin{eqnarray}\label{divqinc} \langle
0|T\psi_+(x)\overline\psi_+(y)|0\rangle=\exp\left\{-\frac{4\pi\,i}{a-1}
\int\!\!
\frac{dk}{(2\pi)^2}\;\frac{1-e^{-ik{\cdotp}(x-y)}}{k^2}\right\}\;
G^{\prime}_+(x-y).\quad \end{eqnarray} The remaining functional
integration in $G^{\prime}_+(x-y)$ involves the following terms
\begin{eqnarray}\label{selfqq} G^{\prime}_+(x-y)& =&\!\!\! N'\int\!\!
d\phi \,d\psi
d\overline\psi\;\;\psi_+(x)\overline\psi_+(y)\;\exp\left[i\int\!\!
dz\;\overline\psi(i\partial\!\!\!\slash-\tilde
\partial\!\!\!\slash\phi\,P_+)\psi\right]\\ & &
\hspace{-1cm}\times\exp\left\{i\int\!\! dz\left[
\frac{1}{2e^2}\phi\square(\square+\frac{e^2}{4\pi(a-1)})\phi+
\frac{1}{a-1}\phi(z)j(z,x,y)\right]\right\}.\nonumber
\end{eqnarray} From (\ref{selfqq}) we easily see that the 1-loop
contribution to the fermion self-energy is finite, as well as the
contribution of the other loops. The ultraviolet divergence is
entirely due to the longitudinal component of the photon, the
$\rho$ field.
\section{Conclusions}

We demonstrated the completely non-perturbative origin of
divergences that occur in fermionic correlation functions in two
dimensional massless Quantum Electrodynamics. This has been done
explicitly, either by summing (wherever it was possible to do it)
the perturbative series or by giving arguments that showed the
finiteness of individual terms. It resulted clear that the
divergences are a consequence of the lack of gauge invariance (at
least in intermediate steps) and that their structure is largely
independent of the fact that the anomaly is a genuine one (as is
the case for the chiral Schwinger model) or an artifact of
regularization (as in the Schwinger model considered under a
general regularization, not necessarily preserving gauge
invariance). The only difference between the two cases is that, in
the second case, the divergences could be circumvented by choosing
a regularization that conducted to $a=1$ (preservation of gauge
invariance). Apart from this fact (which has its justification
only on simplicity, not reflecting any fundamental principle of
quantum field theory) there is no reason for choosing one or
another value for $a$ as, in the end, the effective action (the
one obtained after integration over the fermions and the
Wess-Zumino fields (GIF) or the longitudinal part of the gauge
field (GNIF)) is gauge invariant \cite{HaradaT}.

Moreover, we performed this demonstration both in GIF and GNIF and
found equal results. Although this may seem to be no surprise, as
we were merely effecting the same integral by different means, it
contradicts what is said by Jian-Ge, Qing-Hai and Yao-Yang in
\cite{chineses-JP}. In their paper, the authors find that, when
they add the Wess-Zumino term to the original action, there is no
divergence in fermionic Green functions, as opposed to what they
obtain in its absence, and they justify this exhibiting a
different ultraviolet behaviour of the photon propagator in the
two approaches. In fact, they missed a crucial point in their
paper: they added the Wess-Zumino term {\it before} the
introduction of the external sources, what is {\it wrong}. If one
does this, one does not obtain the coupling between the fermions
and the Wess-Zumino fields, and it is no surprise if no divergence
appears. Also, they obtained two different photon propagators, one
in the GIF and other in the GNIF, because they lost an additional
coupling term between the photon and the Wess-Zumino field, that
comes from the Faddeev-Popov procedure, and that is crucial for
the final expressions for the photon propagators. The correct
expression for the generating functional in GIF is given by our
equation (\ref{gfunt}).

We would like to remember that the generating functional has to be
a functional of something, so the starting expression has to
include the external sources. This is just one instance where this
kind of mistake can conduct one to completely wrong results, and
it is not usually noticed (for an example of the crucial role
played by external sources see \cite{Tiao-Teresa}). If the
external sources are carefully considered from the beginning, one
finds exactly the same results in both formalisms.

The physical interpretation of this new type of divergence is
still unknown for us. The origin of conventional ultraviolet
divergences can be traced back to the requirement of relativistic
covariance and {\it non-triviality} of the field theory under
investigation \cite{strocchi-LNP}. This prevents a good definition
for the field as an operator for all points of the support,
introducing divergences when its powers appear. They manifest
themselves in the diagonal of Green's functions, thus allowing
themselves to be renormalized through the well known ambiguity of
these diagonals under local integrated polinomials in the fields
(counterterms). In the new situation, although the cure may be
similar (but significantly different \cite{tiao-rodolfo}), the
disease may not be the same. The new divergence that appeared
multiplies an effective (and finite) two-point fermionic function.
We called it {\it ultraviolet} just because of its form (look at
the divergent integral appearing in (\ref{petfer22}), for example)
but not because of its origin. After all, it has its origins in an
integration over a quadratic portion of the action, which would
contribute with linear terms in the equations of motion, not
usually associated to divergences (they do not involve products of
operators in the same point). There are evident connections
between this divergence and the lack of gauge invariance, but they
do not help to clarify the situation, as opposed to what was said
above, about conventional UV divergences. Further investigation on
this question is being done, and will be reported elsewhere.

\end{document}